\documentclass[epsfig,12pt]{article}
\usepackage{epsfig}
\usepackage{graphicx}
\usepackage{amssymb}
\usepackage{cite}
\usepackage{amsmath}
\usepackage{hyperref}

\begin{document}
\unitlength = 1mm

\def\de{\partial}
\def\Tr{ \hbox{\rm Tr}}
\def\const{\hbox {\rm const.}}  
\def\o{\over}
\def\im{\hbox{\rm Im}}
\def\re{\hbox{\rm Re}}
\def\bra{\langle}\def\ket{\rangle}
\def\Arg{\hbox {\rm Arg}}
\def\Re{\hbox {\rm Re}}
\def\Im{\hbox {\rm Im}}
\def\diag{\hbox{\rm diag}}


\def\QATOPD#1#2#3#4{{#3 \atopwithdelims#1#2 #4}}
\def\stackunder#1#2{\mathrel{\mathop{#2}\limits_{#1}}}
\def\stackreb#1#2{\mathrel{\mathop{#2}\limits_{#1}}}
\def\Tr{{\rm Tr}}
\def\res{{\rm res}}
\def\Bf#1{\mbox{\boldmath $#1$}}
\def\balpha{{\Bf\alpha}}
\def\bbeta{{\Bf\beta}}
\def\bgamma{{\Bf\gamma}}
\def\bnu{{\Bf\nu}}
\def\bmu{{\Bf\mu}}
\def\bphi{{\Bf\phi}}
\def\bPhi{{\Bf\Phi}}
\def\bomega{{\Bf\omega}}
\def\blambda{{\Bf\lambda}}
\def\brho{{\Bf\rho}}
\def\bsigma{{\bfit\sigma}}
\def\bxi{{\Bf\xi}}
\def\bbeta{{\Bf\eta}}
\def\d{\partial}
\def\der#1#2{\frac{\d{#1}}{\d{#2}}}
\def\Im{{\rm Im}}
\def\Re{{\rm Re}}
\def\rank{{\rm rank}}
\def\diag{{\rm diag}}
\def\2{{1\over 2}}
\def\ntwo{${\mathcal N}=2\;$}
\def\nfour{${\mathcal N}=4\;$}
\def\none{${\mathcal N}=1\;$}
\def\ntwot{${\mathcal N}=(2,2)\;$}
\def\ntwoo{${\mathcal N}=(0,2)\;$}
\def\x{\stackrel{\otimes}{,}}

\newcommand{\cpn}{CP$(N-1)\;$}
\newcommand{\wcpn}{wCP$_{N,\widetilde{N}}(N_f-1)\;$}
\newcommand{\wcpd}{wCP$_{\widetilde{N},N}(N_f-1)\;$}
\newcommand{\vp}{\varphi}
\newcommand{\pt}{\partial}
\newcommand{\tN}{\widetilde{N}}
\newcommand{\ve}{\varepsilon}
\renewcommand{\theequation}{\thesection.\arabic{equation}}

\newcommand{\sun}{SU$(N)\;$}

\setcounter{footnote}0

\vfill

\begin{titlepage}

\begin{flushright}
FTPI-MINN-18/04, UMN-TH-3712/18\\
\end{flushright}

\begin{center}
{  \Large \bf  
Supersymmetrizing the GSY Soliton
\\[2mm]
}

\vspace{5mm}
\vspace{1mm}

{\large \bf   E. Ireson$^{\,a}$, M.~Shifman$^{\,a}$, and \bf A.~Yung$^{\,\,a,b,c}$}
\end {center}

\begin{center}

$^a${\it  William I. Fine Theoretical Physics Institute,
University of Minnesota,
Minneapolis, MN 55455}\\
$^{b}${\it National Research Center ``Kurchatov Institute'', 
Petersburg Nuclear Physics Institute, Gatchina, St. Petersburg
188300, Russia}\\
$^{c}${\it  St. Petersburg State University,
 Universitetskaya nab., St. Petersburg 199034, Russia}
\end{center}
\vspace{1cm}
\begin{center}
{\large\bf Abstract}
\end{center}

We supersymmetrise the Hopfion studied in \cite{GorskySY}. This soliton represents a closed
semilocal vortex string in U(1) gauge theory. It carries nonzero Hopf number due to
the additional winding of a phase modulus as one moves along the closed string. We study this solution in \ntwo 
supersymmetric QED with two flavours. As a preliminary exercise we compactify one space dimension and consider a straight
vortex with periodic boundary conditions. It turns out to be 1/2-BPS saturated. An additional winding along the string can be introduced and it does not spoil the BPS nature of the object. Next, we consider a ring-like vortex in a non-compact space and show that the circumference of
the ring $L$ can be stabilised once the previously mentioned winding along the string is introduced. Of course the ring-like
vortex is not BPS but its energy becomes close to the BPS bound if $L$ is large, which can be guaranteed in the case that we have a large value
of the angular momentum $J$. Thus we arrive at the concept of asymptotically BPS-saturated solitons. 
BPS saturation is achieved in the limit $J\to \infty$.
%
%
%
%
%
\end{titlepage}
\newpage


\section{Introduction }

\setcounter{equation}{0}

Several years ago Gorsky, Shifman and Yung considered a Hopf-type soliton, i.e. with two different types of windings \cite{GorskySY}.
This soliton was explicitly constructed as a closed Abelian 
semilocal vortex string in QED with two flavours and a special type of potential. Although the ``bulk'' model 
in \cite{GorskySY} was nonsupersymmetric, it was
inspired by the previous studies of supersymmetric QED (SQED). 

\begin{figure}[h]
\centering
\includegraphics[width=0.7\textwidth]{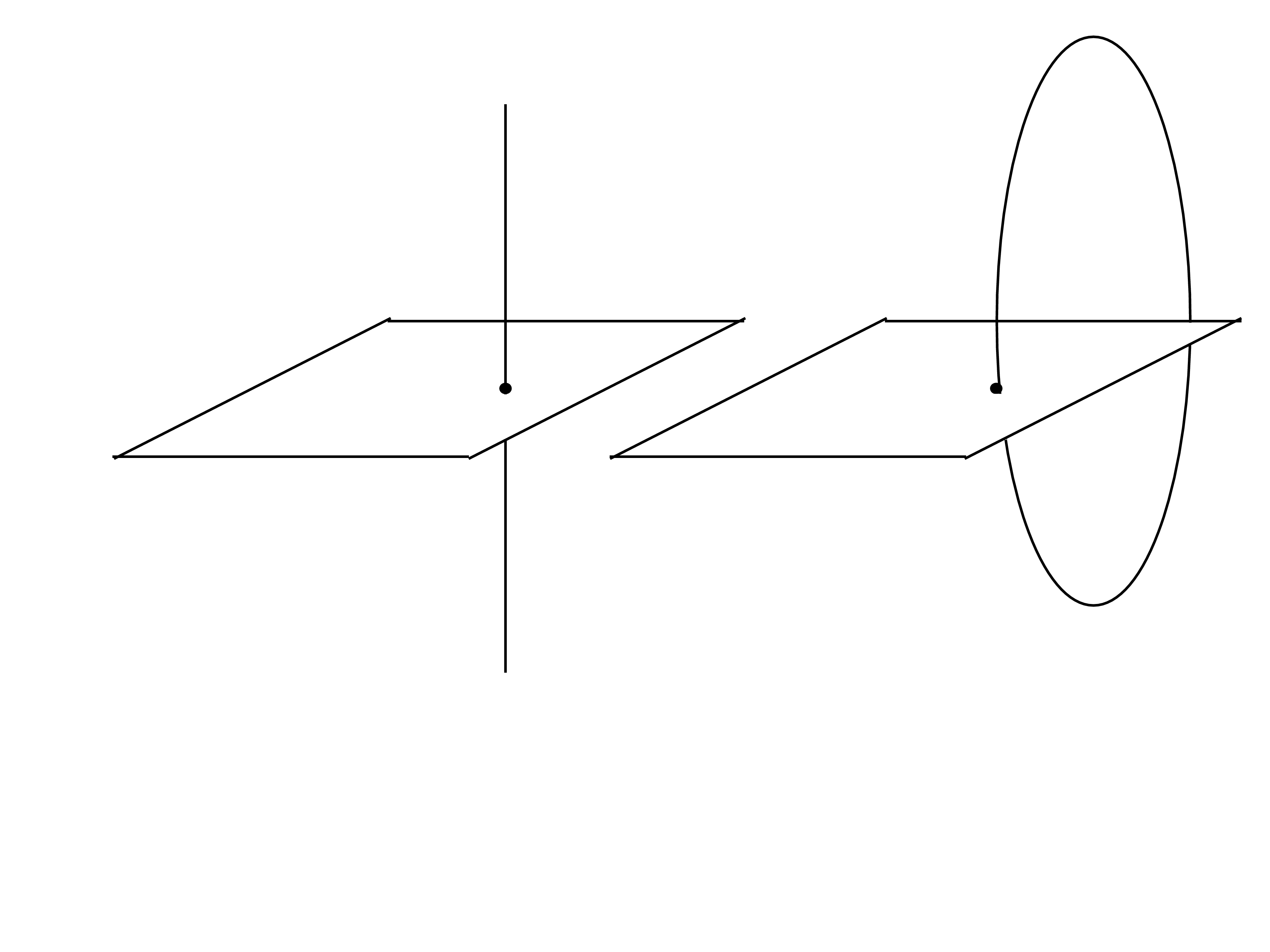}
\caption{\small  Linear and circular vortex ``thin''  strings.}
\label{pic}
\end{figure}

In this paper we present a supersymmetric version of the model considered 
in \cite{GorskySY} using the framework of \cite{SYone}. In the latter, the linear  vortex string  is a BPS saturated object of great interest, since the emerging world-sheet sigma model arising from its quantisation
is conformal. In this paper, we construct a closed circular vortex string which satisfies the condition
\begin{equation}
M^2 = J \times 8\pi T\, \,\, {\rm at}\,\,  J\gg 1\,,
\label{th1}
\end{equation}
where
\begin{equation}
T= 2\pi\xi
\label{th2}
\end{equation}
is the exact string tension. The parameter $\xi$ is the Fayet-Iliopoulos coefficient, to be defined below. This string tension is produced by a winding of certain fields in the  plane transverse to the string (see Fig. \ref{pic}) while $J$, {\em the angular momentum} of the closed string configuration, is generated by the winding of fields around the string itself. Equation (\ref{th1}) is valid to the leading order in $J$. Corrections run in powers of $J^{-1}$ and presumably vanish in the strong coupling regime of infinitely heavy Higgsed gauge bosons, see \cite{SYone}. The issue of subleading corrections will be discussed separately.

First we will discuss the internal structure of the linear BPS string in SQED with two flavours in conjunction with the appropriate superalgebra.  We will then proceed to add the second winding, and then to make the string circular. In the linear version, which serves as an auxiliary exercise in the construction of the closed Hopf-like string, our results are exact.

Closed strings stabilized by a large angular momentum were discussed in the past in the framework of string theory, see e.g.  \cite{st1,st2}.
In \cite {mmt} rotating strings in AdS$_5\times$S$^5$ with $SO(6)$ angular momentum were shown to become asymptotically BPS-saturated in the
limit of infinite momentum: they preserved 1/8-supersymmetry.

The organization of the paper is as follows. In Sec. \ref{pac} we formulate our model and consider first 
a straight BPS string in a compact space with two windings and periodic boundary condition, then, we will develop the main features of an almost BPS-saturated ring-like vortex. In Sec. \ref{analysis} we consider an  explicit solution for a semilocal ring-like vortex. We observe a Bogomolny bound and supersymmetry transformations which produce first order equations. We also relate the mass of the  soliton to the value of the Hopf invariant.

\section{Preliminaries and concepts}
\label{pac}
\setcounter{equation}{0}

\subsection{Model}

The inspiration for this analysis comes from \cite{GorskySY} (see also references therein), 
which will be supersymmetrised. 

The bulk model is ${\cal N}=2$ SQED with a Fayet-Iliopoulos term and two charged flavours,
\begin{eqnarray}
\label{sqedp}
&&
{\mathcal L } = \left\{ \frac{1}{4\, e^2}\int\!{d}^2\theta \, W^2 + {\rm
H.c.}\right\} 
+\left( \int d^2\theta\,\tilde{Q}_A\,(\sqrt{2}\mathcal{A} +m_A)\, Q^A +\text{H.c. }\right)
\nonumber
\\[3mm] 
&& +
\int \!{ d}^4\theta \sum_{A=1,2} \,\bar{\tilde{Q}}_A\, e^{-V}\, \tilde{Q}^A
+
\int \!{d}^4\theta \sum_{A=1,2} \,\bar{Q}_A\, e^{V}\, Q^A 
\nonumber
\\[3mm] 
&& -  \xi  \int\! {d}^4\theta 
\,V(\! x,\theta , \bar\theta ) \, ,
\end{eqnarray}
where  $Q$ and $\tilde{Q}$ are  chiral matter 
superfields with masses $m_A$ and electric charges $\pm 1$, respectively, $A=1,2$ is the flavour index,
$W_\alpha$ is the field strength for the vector superfield $V$,
\begin{equation}
{W}_{\alpha} = \frac{1}{8}\;\bar{D}^2\, D_{\alpha } V =
  i\left( \lambda_{\alpha} + i\theta_{\alpha}D - \theta^{\beta}\, 
F_{\alpha\beta} - 
i\theta^2{\partial}_{\alpha\dot\alpha}\bar{\lambda}^{\dot\alpha} 
\right)\, . 
\label{sgpfst}
\end{equation}
and $\mathcal{A}$ is a chiral superfield containing the extra scalar and fermion components of the $\mathcal{N}=2$ vector multiplet.
The Fayet-Iliopoulos parameter $\xi$ is introduced in (\ref{sqedp}) which is needed to make our construction BPS saturated.
The need for the introduction of a second flavour will become clear shortly. 

After passing to components (in the Wess--Zumino gauge), setting fermions 
to zero, we arrive at the action in 
the following form:
\begin{eqnarray}
S &=&\int d^4 x \left\{
-\frac{1}{4e^2}\, F_{\mu\nu}\,F^{\mu\nu} + \frac1{e^2}
\left|\partial_{\mu}a\right|^2 
-V(q,\tilde{q},a) 
\right.
\nonumber\\[2mm]
&+&\sum_{A=1,2}  \left[ 
{\mathcal D}^\mu\bar{q}_A\, {\mathcal D}_\mu q^A
+
{\mathcal D}^\mu\bar{\tilde q}_A\, {\mathcal D}_\mu \tilde q^A  \right]\Big\}\,.
\label{n1bt}
\end{eqnarray}
Here
$q^A,\,\tilde q^A $ and $a$ are scalar fields belonging to
$Q^A$, $\tilde Q^A$ and $\mathcal{A}$, respectively. 
Covariant derivative is defined as 
\begin{equation}
{\mathcal D}_\mu =\pt_{\mu} -iA_{\mu}.
\end{equation}
The scalar potential is given by the sum of the $D$ and $F$ terms,
\begin{eqnarray}
V(q,\tilde{q},a) &=& \frac{e^2}{2}\,  \left[ \xi - 
\sum_{A=1,2} \left( \bar{q}_A\,q^A -  \tilde q_A \,\bar{\tilde q}^A\right) +
\right]^2 + 2e^2\,\left|\sum_{A=1,2}  \tilde{q}_A\,q^A\right|^2
\nonumber\\[2mm] 
&+&
\sum_{A=1,2} \left\{ \left|(\sqrt{2}a+m_A )q^A\right|^2
+\left|(\sqrt{2}a+m_A)\bar{\tilde{q}}^A
\right|^2 \right\}\,.
 \label{odynone}
\end{eqnarray}
Without loss of generality, we can  assume that the Fayet-Iliopoulos parameter is positive,
\begin{equation}
\xi > 0\,.
\end{equation}
This can always be achieved: if $\xi$ was originally negative, we can make it positive by making a $C$ transformation. 

We should note that only the difference of the electron masses $\Delta m= m_1-m_2$ has a physical meaning, because their sum always can be
turned to zero by  a shift of the complex scalar $a$, a superpartner of the photon.
For a generic choice of $\Delta m$ we have two  isolated vacua in the above theory
with $\langle a \rangle =-\Delta m/2\sqrt{2}$ or $\langle a \rangle =\Delta m/2\sqrt{2}$.
However in the equal mass limit,  
\begin{equation}
\Delta m =0
\label{equalmasses} 
\end{equation}
which we mostly consider below two vacua coalesce and a Higgs branch develops from the common root at 
\begin{equation}
\langle a \rangle = 0.
\label{avev}
\end{equation}

The generic vacuum manifold determined by the constraint $V=0$ is four-dimensional, but we can  reduce it to two dimensions 
by setting the tilded fields to zero, $\tilde Q^A =0$. 
Then the tilded fields will play no role on the  string solution, neither will the scalar $a$, which is given by its VEV in Eq. (\ref{avev}). 
This choice  is self-consistent.

The vacuum manifold is determined by the equation
\begin{equation}
|q^1|^2 + |q^2|^2 = \xi\,,
\label{26}
\end{equation}
with a common phase eaten by the Higgs mechanism. This is a sphere $S_2$. We call it a base of the  
four dimensional Higgs branch. The string can be BPS saturated only if we restrict ourselves to the base
manifold (\ref{26}). String solutions in a generic vacuum with nonzero $\tilde{q}$ are not BPS \cite{EvlY}.

\begin{figure}
\centering
\includegraphics[width=0.7\textwidth]{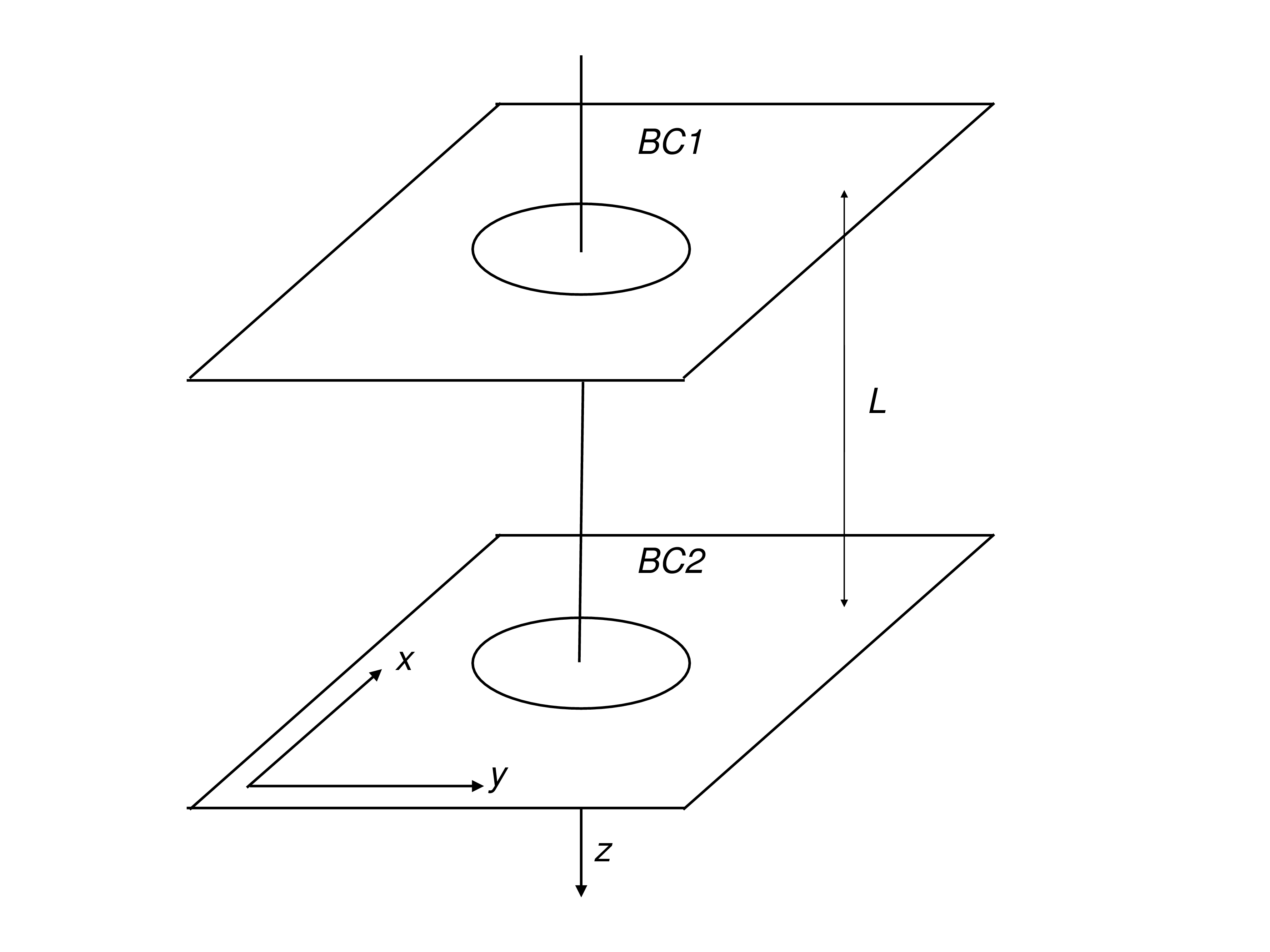}
\caption{\small  Spatial geometry. An intermediate step between the linear string and the circular one
is imposing periodic boundary conditions (BC) in the $z$ direction (with the period $L$), i.e. $BC1=BC2$.   The soliton axis is aligned with the $z$ axis.
The axis $y$ here will correspond $u$ in what follows, see Fig. \ref{halfof}. One can introduce the radial parameter $ r=\sqrt{x^2+y^2}$, see e.g. Eq. (\ref{Ansatz1}).}
\label{fone}
\end{figure}

Using the $SU(2)$ flavour symmetry one can always say that in the 
vacuum (which also means far away from the soliton core which is at the origin  in the $(x,y)$ plane, see Fig. \ref{fone})
\begin{equation}
|q^1|^2 = \xi\,\, {\rm and }\,\, q^2 = 0\,.
\end{equation}
Of course, inside the soliton both fields $q^1$ and $q^2$ can and will appear. Moreover, since $q^1$ will have a winding in the $(x,y)$ plane, it must vanish in the core center. If so, it becomes energetically expedient to develop a non-vanishing value of $q^2$ in the core (see \cite{GorskySY}).

We pause here to make our definition of the linear string ``core'' more precise. In fact, the core has two components. The so-called ``hard'' 
core has the thickness of the order of the inverse mass of the Higgsed photon, $\ell_h \sim (e^2\xi )^{-1/2}$. 
This is similar to the standard ANO string. 
However, as we will see 
shortly (see also \cite{GorskySY}), the existence of the second flavour implies that an additional complex moduli $\rho$ is develops on the string world sheet.
The emergence of $\rho$ is due to Belavin-Polyakov instantons \cite{BP} on the vacuum manifold (\ref{26}).\footnote{These are the Belavin-Polyakov instantons \cite{BP} in the two-dimensional  $O(3)$ sigma model.} The absolute value of $\rho$ plays the role of the string thickness. Outside the hard core
the soliton solution falls off with distance from the center according to a power law, rather than exponentially. Thus, the string at hand is semilocal, the norms of the solution and some zero modes logarithmically diverge, for more details see \cite{GorskySY} and the reviews \cite{Achu,sybook}. In order to make the thickness of the ``soft" core
finite\,\footnote{This will be needed for the construction of the closed string below.} we must introduce an infrared regularization into the theory under consideration. 
The most natural way of the IR regularization is to introduce very small mass difference $\Delta m\sim \mu_{\rm IR}$. We will assume that not only
 $(e^2\xi )/\mu_{\rm IR}^2\gg 1$, but the logarithm $$\log \frac{e^2\xi }{\mu_{\rm IR}^2 } \gg 1$$ too. For more details see \cite{sybook}. 
When the distance from the center in the $\{x,y\}$ plane (Fig. \ref{fone}) exceeds $\mu_{\rm IR}^{-1}$, the power-decaying functions in the solution become exponentially decaying.

The linear string solution {\em per se} has no $z$ dependence. Upon quantization of the moduli, they become $t,\,z$ dependent moduli fields and produce 
a two-dimensional sigma model. For the time being we will consider a linear string of Fig. \ref{pic}a. We will introduce a second further winding, in addition to that inherent to the ANO string. 


Remembering that our final goal is transforming the genuine BPS-saturated linear string  into a circular one, which can be viewed as approximately BPS-saturated in the limit of large $J$, we will take an intermediate step. Note that the circular string cannot be exactly a BPS object, that is, form a short multiplet of supersymmetry, because there is no appropriate global central charge in the superalgebra, but the linear string can. After studying the linear string, we will show that for the closed circular string,
there exists  an approximate lower bound of the Bogomolny type, which we can approach to leading order in our approximation.

\subsection{Linear string and periodic boundary conditions}
\label{lsapbc}

Let us consider the linear BPS string in the geometry of Fig. \ref{fone}, i.e. we lift the requirement of  full $z$ independence of the soliton solution, and impose instead periodic boundary conditions in the $z$ direction.  Our task is to introduce an additional winding of a field along the $z$ axis, which will result in a Hopfion-type field configuration (that is, one with two different types of windings). The field $q^1$ cannot wind along the $z$ direction, since this will produce an infinite amount of energy in the $(x,y)$ plane. On the other hand, nothing prevents $q^2$ from winding since this field falls off at infinity in the $(x,y)$ plane. The periodicity condition then naturally requires that the $z$-dependence is periodic,\footnote{In this section the interval of periodicity is denoted by $L$. In the subsequent sections $L$ will be used for circumference of the closed circular string. We assume that $L\gg \left(\mu_{\rm IR}\right)^{-1}$.  } that is, that the field has a winding number:
\begin{equation}
q^2 (x,y,z,t) = q^2 (x,y) \,e^{i\alpha (z,t)}\,,\, \qquad  \alpha (z,t) = \frac{2\pi k}{L} \, (z\pm t) \,.
\label{fr210}
\end{equation}

The solution (\ref{fr210}) represents left- and right-moving plane waves propagating in the $\pm z$ direction inside the vortex string. One readily calculates the momentum carried by this 
wave,
\begin{equation}
p^z = \int\, d^3 x \, \Theta^{tz}
\end{equation}
where $\Theta^{\mu\nu}$ is the energy-momentum tensor. Since $q^2$ is the only part of the configuration that is dependent on $t,z$ we have 
\begin{equation}
\int \, d^3x\, \Theta^{tz} = \int\,d^3 x \left( D^0 \bar{q}^2 \, D^3 q^2 +  D^3 \bar{q}^2 \, D^0 q^2\right) \sim \frac{k^2}{L} \int\, d^2 x_\perp \left|q^2\right|^2\,.
\label{211t}
\end{equation}
As one could expect, $p^z$ is simply proportional to $L^{-1}$.

The expression for the corresponding energy is 
\begin{equation}
\int \, d^3x\, \Theta^{tt} = \int\,d^3 x \, 2 \left( D^0 \bar{q}^2 \, D^0 q^2 \right) \sim \frac{k^2}{L} \int\, d^2 x_\perp \left|q^2\right|^2\,.
\end{equation}
The energy is equal to the absolute value of the momentum in the $z$ direction, as is obvious, of course, from Eq. (\ref{fr210}).

\subsection{Superalgebra}
\label{sa}

For simplicity we restrict ourselves to \none part of the superalgebra in this section, see Sec. 3.5
for \ntwo analysis.

The \none subalgebra obeyed by the supercharges in the case at hand takes the form
\begin{eqnarray}
\{Q_\alpha\, , \bar Q_{\dot\alpha}\}
 =  2 P_{\alpha\dot\alpha}
+2 Z_{\alpha\dot\alpha}  
\equiv
2\left( P_\mu + Z_\mu \right)
\left(\sigma^\mu \right)_{\alpha\dot\alpha}\, ,
\label{bsa}
\end{eqnarray}
where $P_\mu$ is the momentum operator, and
\begin{equation}
Z_\mu = \,\xi\,  \int { d}^3 x\, \epsilon_{0\mu\nu\rho}
\left( \partial^\nu A^\rho \right) + ...
\end{equation}
is the string ``central charge''  \cite{GorskyS2} (CC in what follows, in application to vortex strings referred to as brane charges, see \cite{Komargodski}).
 In the case depicted in Fig. \ref{fone} it has only one non-zero component which can be written as
\begin{equation}
Z= -Z_3 = - L\xi\int d^2x \, B
\end{equation}
where
\begin{equation}
B=
\frac{\partial A_y}{\partial x} - \frac{\partial A_x}{\partial y}\,
\end{equation}
i.e. the $z$ component of the magnetic field. In the rest frame in the $(x,y)$ plane, we choose $P_{1,2}=0$ and denote $P_3 \equiv - p$. This is the momentum carried by the field $q^2$, which is a massless mode, therefore does not vanish in any frame. We also note that $Z_\mu$ is aligned with $\vec p_z$. 

In this case, the superalgebra (\ref{bsa}) reduces to
\begin{equation}
\{Q , \bar Q\} =2 \left(
\begin{array}{cc}
E - p - L\xi \int d^2x \, B & 0\\[3mm]
0 & E +p + L\xi \int d^2x \, B
\end{array}
\right)
\end{equation}
The general condition of the BPS saturation is
\begin{equation}
E  = p + L\xi \int d^2x B\,;
\label{318}
\end{equation}
for which $Q_1$ and $\bar{Q}_{\dot 1}$ will annihilate the soliton, while  $Q_2$ and $\bar{Q}_{\dot 2}$ will act nontrivially on the solution, producing fermion zero modes.

As is well-known, the integral $\int d^2x B$ is quantised on the solution at hand \cite{sybook}, 
\begin{equation} \int d^2x B =2\pi n
\label{quantn}
\end{equation}  
where $n$ is the integer winding number in the $(x,y)$ plane. For the minimal string we take $n=1$; in what follows we will assume it from now on for simplicity.

The linear string with periodic boundary conditions in the $z$ direction has two windings, and is 1/2 BPS saturated and topologically stable. 
One cannot expect the closed circular string with the double winding to be exactly BPS saturated. However, it is intuitively clear that as the circumference of the circular string 
becomes much larger than its transverse size, i.e. at
 $L\gg  \left(\mu_{\rm IR}\right)^{-1}$, it approaches the BPS bound, and at $L\to\infty$ there is no difference between our pedagogical example and the actual  circular vortex string.

\subsection{Outlining how to make a circular vortex string}

One can make a circular string by bending a linear one, see Fig. \ref{pic}. For self-consistency, we need to do this in such a way that diametrically-opposite points of the core do not overlap significantly, as in general the semi-local solution described previously is not a solution to a linear system of equations. This is especially important given that the fields at hand have a power-law decay rather than an exponential one. It must be assumed therefore that the length of the circular loop $L$ is a very large scale of the problem, in particular compared to the size of the vortex core, so that the string looks long and thin, away from the so-called ``thick string regime." The winding of the $q^2$ field generates angular momentum, in integer units, which contributes to the mass of the object: in the rest frame

\begin{equation}
M= \frac{2\pi J}{L} + 2\pi L\xi\,.
\label{219}
\end{equation}
Comparison with the superalgebra (\ref{318}) is crucial in order to determine the coefficient in front of the
$1/L$ part in (\ref{219}) in terms of the quantum number $J = R |p|$ where $R$ is the radius of the circle in Fig. \ref{pic}\,b. The occurrence of the $1/L$ term was
known previously (see \cite{GorskySY} and references therein) but the coefficient in front of $1/L$ was obtained
in terms of an integral depending on details of the particular solution. The formula (\ref{219}) becomes exact in the limit  $J\to\infty$. At finite values of $J$ corrections
in powers of $1/J$ exists, see Sec. \ref{subs-error}.

Comparing Eqs. (\ref{211t}) and (\ref{219}) we see that 
\begin{equation}
J \sim k^2\, |\rho|^2\, \xi \left( \log \, \frac{1}{\mu_{\rm IR}^2|\rho|^2}\right).
\end{equation}
Provided $|\rho|\mu_{\rm IR}$ is small, this naturally makes the angular momentum a large quantity.\footnote{An alternative is to assume non-minimal winding in the $xy$ plane, $n\gg 1$.} 
The above strong inequality justifies our approximation. We should make a reservation, however. Unlike the linear string with the double winding
the circular one can presumably decay through tunneling with the amplitude $\sim \exp (-J)$.

We also note that classically the minimization with respect to the string transverse
size $|\rho|$ would give $|\rho| =0$ for this (almost) BPS case, as detailed in \cite{GorskySY}. However results of 
\cite{SYone} show that quantum effects at strong coupling stabilise $|\rho|$ even in \ntwo supersymmetric
theory.

\vspace{2mm}

The minimum of the right-hand side of (\ref{219}) is achieved at
\begin{equation}
L_* =\sqrt{ \frac{J}{\xi}}
\label{220}
\end{equation}
guaranteeing that $L_*\, \mu_{\rm IR} \gg 1$.
Thus, so long as the total angular momentum is large enough, our solution is self-consistent. 

The value of the right-hand side (\ref{219})  at the minimum  is $4\pi \sqrt{J\xi}$ implying 
\begin{equation}
M^2=8\pi \,TJ
\end{equation}
where $T=2\pi\xi\int d^2x \, B$ is the string tension. Both $T$ and $J$ are proportional to integers, characterising two different types of windings that the fields composing the solution can bear. In this scenario, we will then be able to show that our solution saturates a Hopfian type topological invariant
\begin{equation}	
{\cal H} = \frac{1}{4\pi^2} \int d^3 x \left( { A}_\alpha F_{\mu\nu}\right)  \varepsilon^{\alpha\mu\nu}\,,
\end{equation}
as we will see in Section \ref{subs-hopf}.

Below a more detailed study of the circular string in SQED is presented.

\section{Detailed analysis}
\label{analysis}
\setcounter{equation}{0}

\subsection{The action in cylindrical coordinates}

In order to analyse the field configuration generated by the toroidal soliton, it is preferable to exploit as best we can its symmetries, in this case its invariance under rotations. For this purpose we will employ cylindrical coordinates from the get-go, this will simplify our task when writing the relevant semi-local {\em Ansatz}. The standard set of cylindrical coordinates have a disadvantage, the radial coordinate is bounded below by $0$, a specificity to which we will have to pay attention. We assume that none of the fields, save for some phase dependence in $q^2$, depend on the angular variable, thus, we need only formulate an {\em Ansatz} for the fields in one half of a transverse slice of the torus, as illustrated below.

\begin{figure}[!h]
	\centering
	\includegraphics[width=0.8\textwidth]{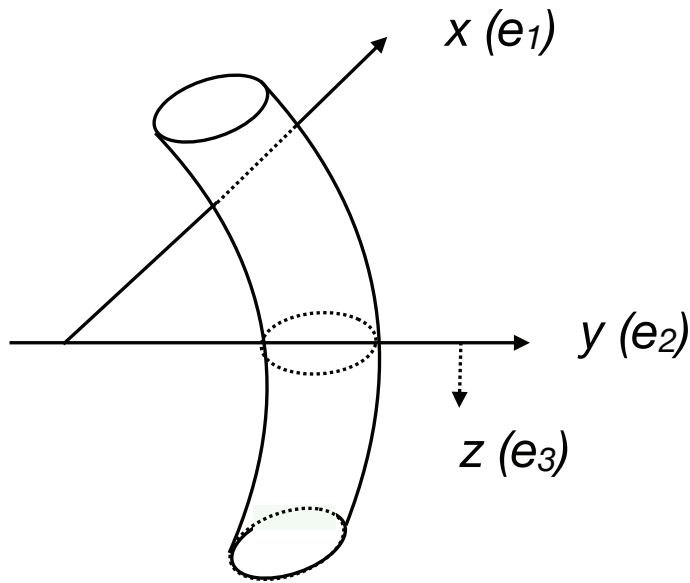}
	\caption{Part of the toroidal configuration. The usual semi-local vortex is inserted at $u=R$, $x=0$. The $z$ axis is perpendicular to the 
	given plane.}
	\label{halfof}
\end{figure}

The torus is supposed symmetric under rotations around the axis $x$. We introduce a set polar coordinates $(u,\theta)$ to parametrise respectively radial motion away from the axis, and circular motion around it. These new coordinates fulfill the role of $(y,z)$ in the straight string case, as shown in Fig. \ref{halfof}. We employ the vierbein formalism, i.e. all objects with space-time indices will be expressed in a local Lorentz basis, which we choose to consist of the following vectors:

\begin{equation}
e^0=dt\,\,,\,\,e^1=dx\,\,,\,\,e^2=du\,\,,\,\,e^3=ud\theta\label{vierbein}
\end{equation}

Numbered indices will from now on correspond to components of objects in this local Lorentz basis, while world indices will be denoted with the letter corresponding to the coordinate.

Importantly, because we are in geometrically flat space, simply using curvilinear coordinates, making the action and the equations of motion covariant is the only step we need to perform in order to have the complete Lagrangian. Non-minimal coupling to gravity such as $R\phi^2$ terms all vanish when the Riemann tensor vanishes. This means that the action expressed in Eq.(\ref{n1bt}) is still formally the right one, so long as every derivative now becomes spacetime-covariant and that the integration measure changes from $dt\,dx\,dy\,dz$ to $dt\,u\,du\,dx\,d\theta$. At this point we can attempt to show that the energy is bounded below by performing Bogomoln'yi completion: let us assume $F_{12}>0$, we then write for the scalar sector

\begin{align}
&(D_1 q)^\dagger (D_1 q) + (D_2 q)^\dagger (D_2 q)\nonumber\\[2mm]
=&(D_1 q - iD_2q)^\dagger (D_1 q - i D_2 q) +i \left( (D_1 q)^\dagger (D_2 q) - (D_2 q)^\dagger (D_1 q)\right) .
\end{align}

The first term is positive-definite, the second one simplifies considerably after integration by parts, which is not altogether trivial in this case as the metric has a non-vanishing determinant. The expressions above are multiplied by $\sqrt{\det g}$ before being integrated, derivatives thereof occur when performing integration by parts. For the second term in the above expression, performing this operation we get

\begin{align}
&\frac{i}{\sqrt{\det g}} q^\dagger \left(D_2^\dagger \sqrt{\det g} D_1 - D_1^\dagger \sqrt{\det g} D_2 \right) q +\text{c.c.} \nonumber	\\[2mm]
=& F_{12}q^\dagger q + i\frac{\partial_u \sqrt{\det g}}{\sqrt{\det g}} q^\dagger\overleftrightarrow{D}_x q - i\frac{
	\partial_x \sqrt{\det g}}{\sqrt{\det g}} q^\dagger\overleftrightarrow{D}_u q \,.
\label{bogo}
\end{align}

The first term in this expression is the result of $\left[D_1,D_2\right]q^\dagger q$ and is the usual expression one gets in Cartesian coordinates. The extra terms are new to our setup: thankfully they simplify considerably\,\footnote{The denominators $\sqrt{\det g}$ in (\ref{bogo}) are cancelled by $\sqrt{\det g}$ in the integration measure. } given that $\sqrt{\det g}=|u|$, so that $\partial_u \sqrt{\det g} = \text{sgn}(u)$, which comes to multiply 
the term $q^\dagger{D}_1 q$.  There is no reason for this term to vanish for generic field configurations, and indeed it does not for correctly-chosen ones, but its contribution is vanishingly small for configurations centered far from the origin around a circle of large radius $R$. Similarly, the integration by parts procedure generates boundary terms
\begin{equation}
\left[\sqrt{\det{g}}\,q^\dagger D_i q\right]_{u=0}^{\infty},
\end{equation}
which need not vanish: certainly the matter currents are expected to decay at infinity on physical grounds, but on the edge of the radial plane, the above term generates a contribution. Again, assuming that the distribution of the current is centered on a point far away from the origin will make this term vanishingly small. The relative importance of subleading corrections due to this approximation will be dealt with in Section \ref{subs-error}.

The former term in the expression above, like in Cartesian coordinates, comes to complete another part of the Lagrangian
\begin{equation}
	-\frac{1}{2e^2}F_{12}^2 -F_{12}q^\dagger q - \frac{e^2}{2}(q^\dagger q - \xi^2) = -\frac{1}{2}\left[ e F_{12}-\frac{1}{e}(q^\dagger q - \xi)\right]^2 + \xi F_{12},
\end{equation}
which is again a positive-definite part and a remainder term. We integrate it over all of space and combine it with the approximated sub-leading term computed above to find an approximate bound for the energy,

\begin{equation}
	E\geq  \xi\int udu\, dx\, d\theta \, F_{12} + O\left( \frac{1}{R}\right).
\end{equation}

We obtain an approximate lower bound for the energy, valid for large configurations localised away from the origin, which is an approximation we will need to assume several times in the following derivations. 

 Finally, if $F_{12}<0$ we can of course complete the squares with opposite signs and get a similar result in terms of $$\left| \!\int udu\, dx\, d\theta \,F_{12}\right|.$$

If the positive-definite terms we have isolated in this derivation can be made to vanish (perhaps only to leading order in $\frac{1}{R}$) then we will obtain a finite-energy solution whose energy is very close to this topological-looking lower bound. Because this derivation is only approximate, the system will never truly be BPS, but the configurations are nevertheless of interest.

\subsection{An  {\em Ansatz} for the fields}

The low energy limit of our theory (\ref{n1bt}) is the O(3) sigma model on the base of the Higgs branch (we restrict ourselves to the {\em Ansatz} with $\tilde{q}=0$, see (\ref{26}).  
The semilocal string
solution at large distances approaches
  the instanton of the two-dimensional O(3) sigma model lifted in four dimensions, see \cite{Achu,EvlY}
	for details.

The semi-local {\em Ansatz} is an approximate solution to the equations of motion that approach a minimal energy configuration, i.e. BPS saturation, even in the case of the straight infinite string, see \cite{sybook}. 
In its original formulation, we write the  {\em Ansatz} for the straight string thus: where $r$ is the radial distance in the plane of the vortex, we introduce the complex core thickness parameter $\rho$ and 
consider the vortex winding (flux ) number $n=1$,
then we write the two scalars and the gauge field in terms of profile functions $F_1$, $F_2$ and $G$ in the following way
\cite{Achu,EvlY}:

\begin{align}
		q^1(r)&=\sqrt{\xi}\frac{r}{\sqrt{r^2+|\rho|^2}}=F_1(r), \nonumber\\[1mm]
		q^2(r)&=\sqrt{\xi}\frac{\rho}{\sqrt{r^2+|\rho|^2}}e^{-i\theta}=F_2(r)e^{-i\theta}, \nonumber\\[1mm]
		A^i&=\frac{\epsilon^{ij}x_j}{r^2}\,f(r) =\frac{|\rho|^2}{r^2(r^2+|\rho|^2)} \epsilon^{ij}x_j=G(r) \epsilon^{ij}x_j \,,\,\,(i=1,2), \nonumber\\[1mm]
		F_{12} &= -\frac1r f'(r),
	\label{Ansatz1}
\end{align}
where we used the expression for the gauge profile function for the semilocal string
\begin{equation}
f(r) = \frac{|\rho|^2}{r^2+|\rho|^2},
\label{gaugeprofile}
\end{equation}
 while prime denotes derivative with respect to $r$.
Note that we are using a singular gauge, so that there is no overall winding at infinity but both the field $q_2$ and the gauge field have singular behaviour at $r=0$. For the generic flux number $n$ the gauge profile function $f(r)$ satisfies boundary conditions
\begin{equation}
f(0) =n, \qquad f(\infty) =0.
\label{boundaryconditions}
\end{equation}
We mostly restrict ourselves to the case $n=1$.

 We adapt this {\em Ansatz} to a curved string of radius $R$. In our coordinates, we must write
\begin{align}
q^1(x,u)&=F_1(x,u-R)+F_1(x,u+R),\nonumber\\[1mm]
q^2(x,u)&=F_2(x,u-R)e^{-i \arctan\left( \frac{u-R}{x}\right)}+F_2(x,u+R)e^{i \arctan\left( \frac{u+R}{x}\right)}\nonumber\\[1mm]
A_i&=\epsilon_{ij}\left[G(x,u-R)(x^j-R^j)-G(x,u+R)(x^j+R^j)\right] ,
\label{Ansatz2}
\end{align}
where $R^i=(0,R)$.

\begin{figure}[h]
	\centering
	\includegraphics[width=0.49\textwidth]{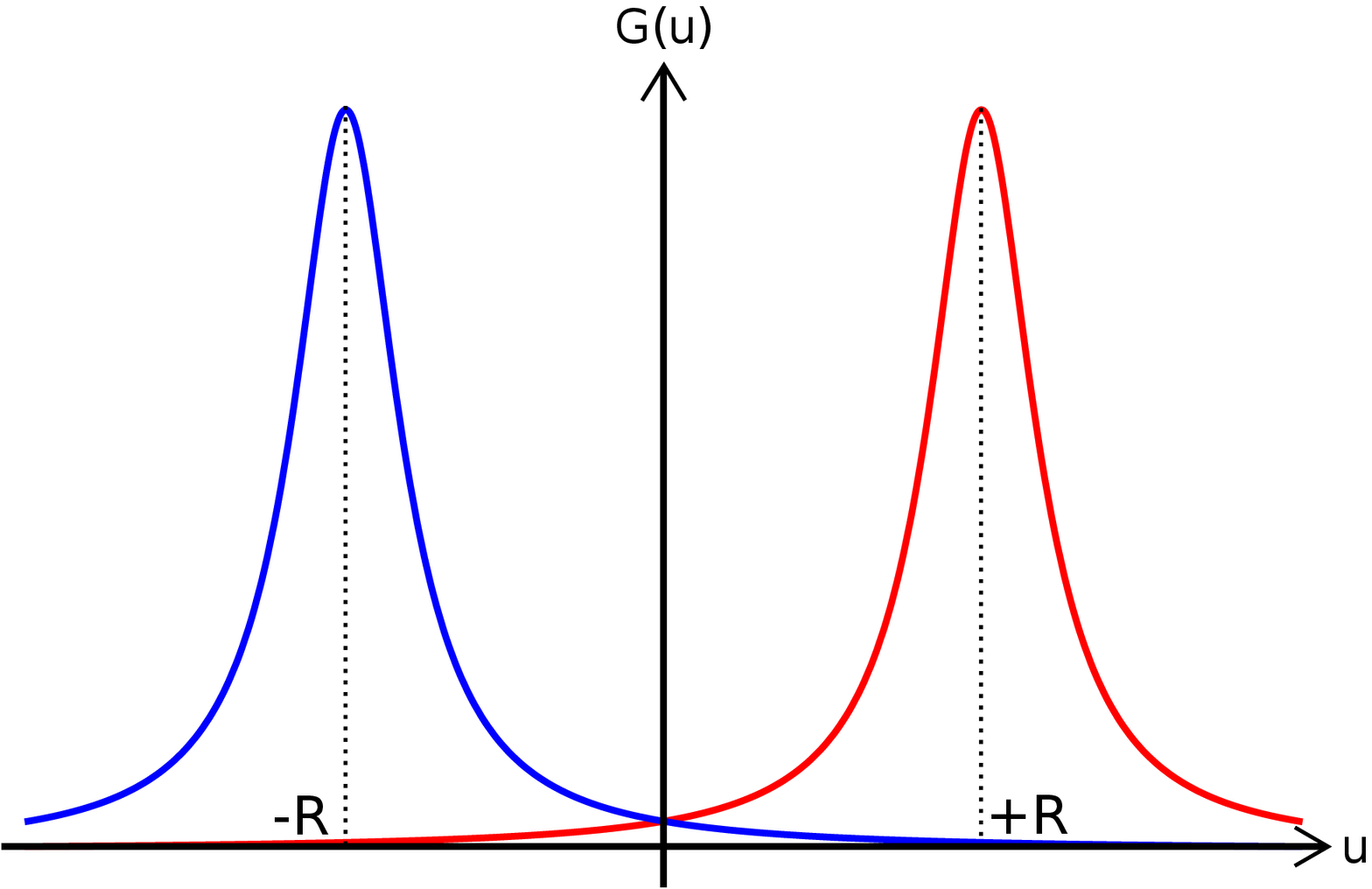}
	\includegraphics[width=0.49\textwidth]{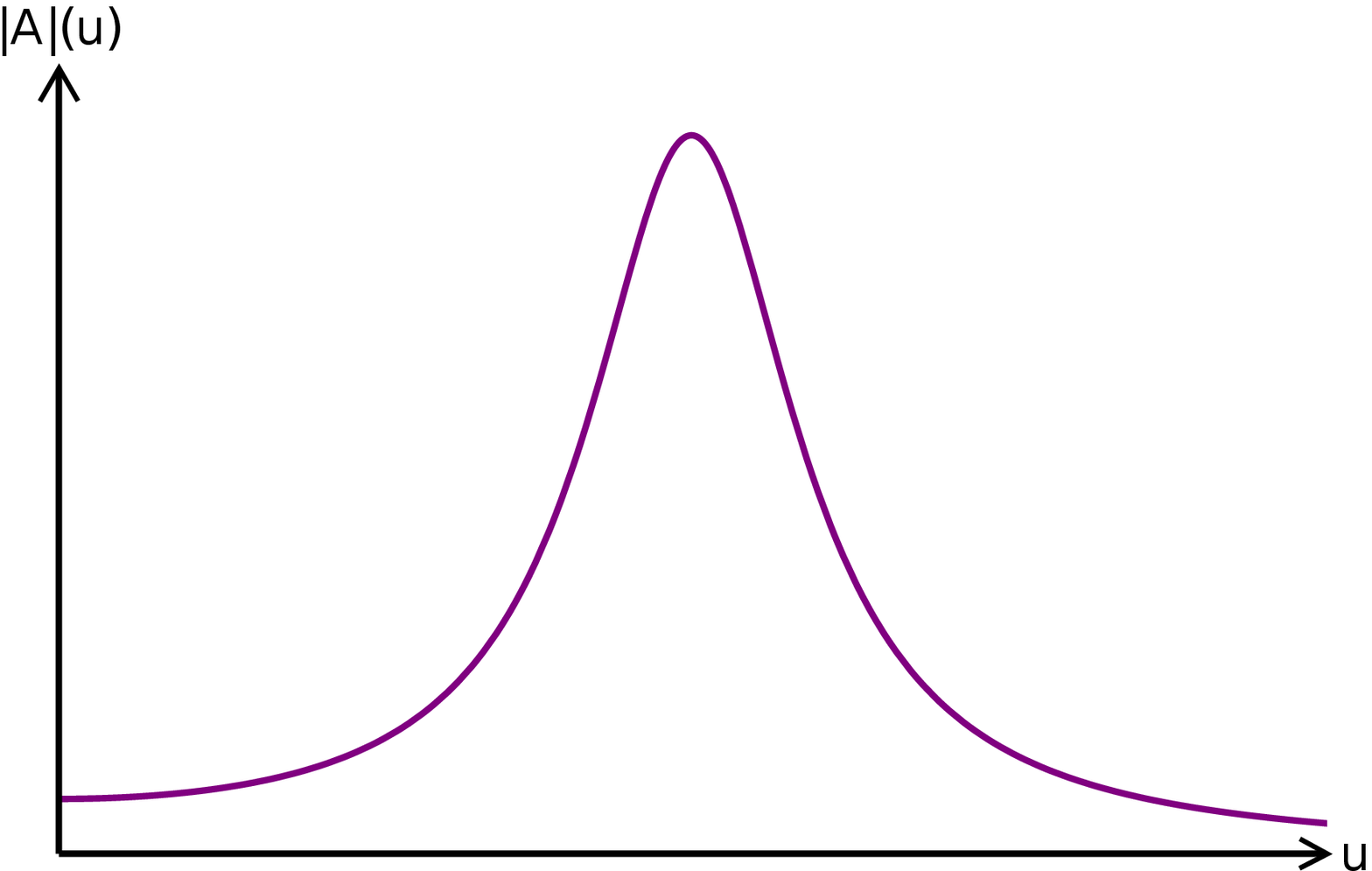}
	\caption{Graphs of the individual vortex and fictitious anti vortex contributions to the total gauge field profile, for fixed $\theta,\,x$. The negative $u$ domain of the first graph is unphysical.}
	\label{fig-profiles}
\end{figure}

This Ansatz is composed of two terms, one due to a vortex centered at $u=R,\, x=0$ and the other being the tail of a fictitious anti-vortex centered at $u=-R,\,x=0$. Though the $u<0$ domain is unphysical, some portion of the tail of this fictitious anti-vortex protrudes into the physical region. The interpretation of this Ansatz is the following: as seen from any particular vortex along the circular string, an anti-vortex is situated diametrically opposite it, on the other side of the torus. Though it is very far away, we should in theory consider that the profiles for these two vortices overlap a little. Figure \ref{fig-profiles} shows a graph of the radial profiles for the gauge field.

From this we can compute the leading term  in the energy bound found earlier,
\begin{equation}
	Z=\int |u|du\, dx\, d\theta\,\,\xi F_{12}\underset{R\gg |\rho|}\approx 2\pi \xi L f(0) =  2\pi n\xi L\,,
\end{equation}
where we used (\ref{Ansatz1}) and (\ref{boundaryconditions}), while $L$ is the total length (circumference) of the vortex core. 

Allowing $L$ to be dynamical, the system clearly favors shrinking as much as it can. At which radius it stabilises is unclear. At the point where the torus becomes of comparable radial size as its cross-sectional width our approximations fail. 

It is possible that there is a stable end-point configuration where the string tension is offset by the energy induced by the overlapping of the vortex configurations, we cannot say.

\subsection{Adding an extra winding}
\label{aaew}

We would like for the configuration to not shrink outside of our initial approximations, for this purpose we introduce an extra winding in the action.

The {\em Ansatz} above in Eq.(\ref{Ansatz1}) can be modified in the following way: the modulus $\rho$, and therefore $q^2$, can have an extra phase, as shown previously. We write
\begin{equation}
\rho=|\rho|e^{i\alpha(t,\theta)}\,,
\label{twist}
\end{equation}
cf. Eq. (\ref{fr210}). The equations of motion read thus
\begin{equation}
	\left( \partial_t^2 + \frac{1}{u^2}\partial_\theta^2\right) \alpha=0\,.
\end{equation}
The appearance of $u$ in the equation is worrisome, we would not like $\alpha$ to appear in the equations of motion of radial fields. We must again use the approximation employed previously, that the length of the string is much larger than local variations of the support of the fields. That is, for all the region where $q^2$ is supported, $u\sim R$. The equation above is then only satisfied up to leading order in $\frac{1}{R}$. At this cost we get
\begin{equation}
	\alpha(t,\theta)= k\left(\frac{t}{R} \pm\theta\right) .
	\label{phase}
\end{equation}
Inserting this in the Hamiltonian, we get that the energy increases by
\begin{equation}
	\Delta E= \int R dudxd\theta |q^2|^2 \left(\frac{2\pi k}{L}\right)^2\,.
\end{equation}
Note again we have replaced $u\rightarrow R$ in the metric determinant. The {\em Ansatz} we have written is nonsensical if we do not perform this operation as it is grossly non-normalisable with this curved metric. Already in flat space the {\em Ansatz} has difficulties, it is logarithmically divergent when attempting to compute its norm. It was argued previously \cite{SVY} that this did not spoil the picture, and indeed in our case it is beneficial. 

We can compute the above integral by introducing a suitable regulator scale $\mu_{\rm IR}$. The process simplifies considerably if we ignore the contributions due to the overlap between the profiles generated by diametrically-opposite points, they lead to higher order terms in the series expansion in $\frac{1}{R}$. We obtain:

\begin{equation}
	\Delta E = \xi \frac{8 k^2\pi^3}{L
	} |\rho|^2  \log \left(\frac{1}{|\rho|\mu_{\rm IR}}\right) \equiv \frac{2\pi J}{L}
\end{equation}
where we have defined the quantity of angular momentum added by the twist
\begin{equation}
	J={4k^2\pi^2} \xi|\rho|^2  \log \left(\frac{1}{|\rho|\mu_{\rm IR}}\right)
	\label{J}
\end{equation}

We recover the form that we guessed previously. This should be a positive quantity, whatever we do. This is very naturally achieved: in order for our assumptions to hold, we must suppose the core size to be very small, at least compared to the scale of far infra-red processes. As $\rho$ is a modulus (in the supersymmetric case), we can pick it to be small in comparison to the IR cutoff. This is actually helpful, it means that the angular momentum contribution to the total energy of the system is actually quite significant, even when we only have one extra winding in the case $k=1$. It is also a self-consistency check for our $\frac{1}{R}$ expansion, despite this term scaling negatively with $R$ it should not be absorbed with our $O(\frac{1}{R})$ terms, since at equilibrium it is expected to contribute as much to the energy as the tension. Let us check this, we can then see again how this counteracts the string tension: the total energy of the system is now

\begin{equation}
	E_{\text{tot.}}= \frac{2\pi J}{L} + 2\pi \xi L + O\left( \frac{1}{R}\right)
\end{equation}

 By extremising the energy over $L$ we find that the system stabilises at a length $L^\star$ given by

\begin{equation}
	L^\star =  2\pi|k| |\rho|\sqrt{\log\left( \frac{1}{|\rho|\mu_{\text{IR}}}\right)}
	\label{L}
\end{equation}
Quite clearly, $L^\star / |\rho| \gg1$ by the arguments above, so our initial {\em Ansatz} is self-consistent. Finally, we can write the mass of the object, placing ourselves in its rest frame, we find
\begin{equation}
	M= 2\pi \xi |k||\rho|  \sqrt{\log \left(\frac{1}{|\rho|\mu_{\text{IR}}}\right)}
	 \label{emin}
\end{equation}
with $J\gg 1$, as advertised.

We can restore the dependence on the flux number $n$ using  the following heuristic argument.
Let us make $n$  toroidal solitons like the one described above in $n$ well separated  planes parallel to each other,  each with magnetic flux $2\pi$. Each of these vortex rings has its own size $\rho_i$, $i=1,...,n$. Now we consider configuration with
\begin{equation}
q^2  = q^2 (x,u) \,e^{i\alpha (t,\theta)}\,,\, \qquad  \alpha(t,\theta)= k\left(\frac{t}{R} \pm\theta\right)  \,.
\label{q2phase}
\end{equation}
This ensures that all $\rho_i$ have the same phase dependence $\alpha(t,\theta)$ determined by a single winding
number $k$. For each of these vortices we introduce the angular momentum $J_i$ given by (\ref{J}) in terms of 
the size $\rho_i$.

Given that the objects are approximately BPS, they generate very little potential energy between them. So, let us adiabatically fuse vortex rings  together. Since both the magnetic flux and angular momentum are
conserved, the fused $n$ multi-soliton   has magnetic  flux number $n$ and total angular momentum $n J $,
where $J$ given by (\ref{J}), and we assume that all $|\rho|$'s are stabilised at the same average value. This gives for the energy of the 
multi-vortex
\begin{equation}
	E_{\text{tot.}}= \frac{2\pi \,J_n}{L} + 2\pi n \,\xi L + O\left( \frac{1}{R}\right),
	\label{vortexenergy}
\end{equation}
where $J_n$ equals to $J$ in (\ref{J}) multiplied by $n$.

Minimizing with respect to $L$ we get the same result (\ref{L}) as for $n=1$ string while the mass of the soliton is given by
\begin{equation}
	M= 2\pi \xi |n||k||\rho|  \sqrt{\log \left(\frac{1}{|\rho|\mu_{\text{IR}}}\right)},
	 \label{emin-n}
\end{equation}
where we assumed that the flux number $n$ could be both positive or negative.

We must perform several other checks on this derivation to ensure it is reasonable. First and foremost, we have included a subleading term in some $\frac{1}{R}$ expansion, but there could be plenty more to add.

\subsection{Estimating the error}\label{subs-error}

For good measure, we must make a note of verifying the self-consistency of the $O\left(\frac{1}{R} \right)$ approximations we have performed. Subleading terms come from four different sources, which should be compared. The first is directly due to the effect of the twist: the term we introduced to stabilise the solution. This is a subleading effect in that it scales with $\frac{\rho}{R}$, but it has a very large numerator to compensate, so we have not neglected it, and shown in the above analysis that the consequences of this choice are self-consistent.

The second comes from neglecting an extra, metric-induced piece of the leftover terms that were produced by performing Bogomoln'yi completion, in Eq.(\ref{bogo}):  $\text{sgn(u)} q^\dagger{D}_x q$. Before substituting the full form of the {\em Ansatz}, we will first only make the assumption that the current component in question is in a toroidal configuration, invariant under rotations in the angle $\theta$, and taking its maximal value on the circle $u=R,x=0$. We make no strong assumptions about the decay of the current or the fields composing it so far, which means points at  angle $\theta$ and $\theta+\pi$ are in theory able to influence each other.  We can thereby write the current with the following substitution:
\begin{equation}
q^\dagger{D}_x q = J_x(u-R,x)+J_x(u+R,x)
\end{equation}
for some regular function $J_x$ that takes its maximum at $(0,0)$. We are tasked to compute

\begin{align}
\Delta E_2=&\iint_{\mathbb{R}^2} du dx \, \text{sgn}(u)\left(  J_x(u-R,x)+J_x(u+R,x)\right)\\ =&\iint_{\mathbb{R}^2} du dx \left( \text{sgn}(u+R)+\text{sgn}(u-R)\right) J_x(u,x)\\
=&2\int_{-\infty}^{\infty} dx \left(-\int_{-\infty}^{-R} J_x(u,x) + \int_{R}^{\infty} J_x(u,x) \right)  .
\end{align}
This expression generically need not vanish, particularly since $J_x$ is expected to not be even in $u$: it is a vector quantity and so is not parity invariant. However, if we assume $R$ to be large, these integrals above should vanish: if $\phi$ is normalisable (or, at worst, with log-divergent norm), then the current $J_x$ should behave this way on either interval above,
\begin{equation}
|J_x(u,x)|=O\left[ \frac{1}{(u^2+x^2)^{3/2}}\right].
\end{equation}
Assuming again that the two leading order contributions need not cancel, after integration this term vanishes at least as $O(\frac{1}{R})$, with only these few assumptions, so that the Bogomoln'yi bound given previously is a good estimate of the lowest available energy of large configurations that peak away from the origin. 

In the case of our {\em Ansatz} it is actually of much lower order: the original semi-local {\em Ansatz} for the straight string generates no net current. The $x$-current generated by $q^1$ is exactly opposite to the current generated by $q^2$,
\begin{equation}
\frac{1}{2i}\left(q^{1\dagger} \overleftrightarrow{D}_x q^1 \right)=  -\frac{\rho^2 u \xi}{\left( \rho^2 + x^2 + u^2\right) ^2} = -\frac{1}{2i}\left(q^{2\dagger} \overleftrightarrow{D}_x q^2 \right)
\end{equation}
Thus, in our case, $J_x$ is only non-zero due to the overlap of the fields generated by diametrically opposite points, which is therefore already a subleading contribution before integration. The scaling arguments above then show that the total contribution after integration must vanish at even higher order than $1/R$. We are therefore justified in ignoring it, as well as the surface term generated via integration by parts, for much the same reasons.

Another source of error comes from the computation of the form of the angular momentum, specifically in computing the normalisation of the radial function $q^2$. We only considered the contributions due to the peaks of the function, assumed widely separated, but there is another piece due to the overlap of the two peaks. This corresponds to the following integral:
\begin{equation}
	\Delta E_3 =\int\,dudx \frac{|\rho|^2 \xi}{\sqrt{(u-R)^2 + x^2 + \rho^2}\sqrt{(u+R)^2 +x^2 + \rho^2}}\,.
\end{equation}
This is a logarithmically divergent integral again, which contains two scales, $\rho$ and $R$. The computation simplifies considerably in the case $\rho\ll R$, which we want to assume throughout. Introducing again an arbitrary mass scale due to regularisation, and up to combinatorial dimensionless constants, this term is proportional to 
\begin{equation}
\Delta E_3\propto	\frac{|\rho|^2\xi}{R} \log\left(\frac1{|\rho|\mu}\right) 
\end{equation}
and can be neglected.

Finally we must investigate the error committed by ignoring the variations of $u$ in the extra twist, we replaced $u\rightarrow R$ and assumed a phase factor that depended only on $(t,\theta)$, which allowed the $t,\theta$ part of the Laplacian to vanish independently of the $(x,u)$ terms. This is not quite correct, with their exact form these terms are
\begin{equation}
	(D_t^2 - \frac{1}{u^2}D_\theta^2)q^2 =\left(  \frac{k}{R}\right) ^2 \left(1-\left( \frac{R}{u} \right)^2\right)q^2\sim \left(  \frac{k}{R}\right) ^2 \left( \frac{\delta u}{
	R}\right)q^2
\end{equation}
where in the last relation we express this term for $u\sim\mp (R+\delta u)$. As a term in the action this is a higher order term in the series that generated the  angular momentum term that we add, so we should not consider it.

Therefore, we believe that the extra term due to the second winding we have added to the theory is indeed the main component, the most influential consequence of the introduction of the extra phase factor, and we conclude that the analysis above is self-consistent.

We have yet to discuss another form of self-consistency. Has this extra mode changed the near-BPS nature of our soliton? To do this we must look at the superalgebra of the theory.

\subsection{Almost-supersymmetric solutions and the central charge}

We know that the soliton at hand is not a true BPS object, so that BPS equations we write for this system are only approximately solved by our version of the semi-local {\em Ansatz}, but their general structure is nevertheless informative. In particular, from our first-principles derivation, there seems to be no correlation between the handedness of the vortex around the core circle and the handedness of the transverse mode in $q^2$. Although strictly speaking we cannot claim our configuration is BPS, inspecting the BPS equations of our Lagrangian at least informs us if we are free to pick the handedness for the transverse modes, to see whether it leads to gross violation of the BPS bound.

 We write the SUSY transformations of the fermionic fields, and impose that they should be zero in such a way as to keep arbitrary some components of the infinitesimal spinor used to parametrise the transformation. We will work with Euclidean conventions for coordinates and $\sigma$ matrices. Let $R$-symmetry indices being denoted abstractly by $f,g\dots$ and in components by Roman numerals $I,II$, we transform each fermionic field with an infinitesimal doublet of spinors $\eta^{\alpha f}$. We assume that the sgaugino (scalar part of the gauge multiplet) vanishes, and introduce the most generic $D$ auxiliary with indices $D^f_g$. 

Finally, we use the following relations expressing the squark $SU(2)$ doublet in terms of the fields $q,\tilde{q}$:
\begin{equation}
q^f=\left(\begin{array}{c}
q\\ 
-i\bar{\tilde{q}}
\end{array}  \right) \,\,\, , \,\,\, \bar{q}_f=\left(\begin{array}{c}
\bar{q}\\ 
i{\tilde{q}}
\end{array}  \right).
\end{equation}
$R$-symmetry indices are raised and lowered with the $\epsilon$ tensor. We can then write
\begin{align}
&\delta_\eta \bar{\psi}_{\dot{\alpha}} = i\sqrt{2}{\eta}^{\alpha f} \bar{\sigma}^\mu_{\dot{\alpha}\alpha }D_\mu \bar{q}_f \,.\\[2mm]
&\delta_\eta \bar{\tilde{\psi}}_{\dot{\alpha}} = i\sqrt{2}\eta^{\alpha f} \bar{\sigma}^\mu_{\dot{\alpha} \alpha}D_\mu q_f\,,\\[2mm]
&\delta_\eta \lambda^f_\alpha = -\eta^{\beta f} (\sigma^\mu\bar{\sigma}^\nu)_{\alpha\beta} F_{\mu\nu} +i\eta^g_\alpha D^f_g\,.
\end{align}

To obtain the untwisted semi-local vortex configuration, we make the choice to preserve $\eta^{1II}$ and $\eta^{2I}$, thus we put $\eta^{2II} = \eta^{1I}=0$. We do not assume any invariances of the fields in any of the coordinates. The above equations produce the following:
\begin{align}
\delta_\eta \bar{\psi}_{\dot{\alpha}} &= i\sqrt{2}\eta^{1II} \bar{\sigma}^\mu_{\dot{\alpha }1}D_\mu\bar{q}_{II}  + i\sqrt{2}\bar{\eta}^{2I} \bar{\sigma}^\mu_{\dot{\alpha}2 }D_\mu \bar{q}_{I}\ \\[1mm]
&=i\sqrt{2}\left(\begin{array}{c}
\eta^{1II}(D_0+iD_3)\bar{q}_{II} +i\eta^{2I} (D_1 - i D_2)\bar{q}_{I}\\ 
\\
i\eta^{1II} (D_1 + i D_2)\bar{q}_{II} + \eta^{2I}(D_0-i D_3)\bar{q}_{I}
\end{array}  \right)\\[1mm]
\nonumber\\
&=i\sqrt{2}\left(\begin{array}{c}
i \eta^{2I} (D_1 - i D_2)\bar{q}\\ 
\\
\eta^{2I}(D_0-iD_3)\bar{q}
\end{array}  \right),
\label{transfo1}
\end{align}
and
\begin{align}
\delta_\eta \bar{\tilde{\psi}}_{\dot{\alpha}} &=i\sqrt{2}\eta^{1II} \bar{\sigma}^\mu_{1 \dot{\alpha} }D_\mu q_{II}  +i\sqrt{2}\eta^{2I} \bar{\sigma}^\mu_{2 \dot{\alpha} }D_\mu q_I 
\\[2mm]
&=i\sqrt{2}\left(\begin{array}{c}
\eta^{1II}(D_0+i D_3)q_{II} +i \eta^{2I} (D_1 - i D_2)q_{I}\\ 
\\
i\eta^{1II} (D_1 + i D_2)q_{II}+\eta^{2I}(D_0-iD_3)q_{I}
\end{array}  \right)\\
\nonumber\\
&=i\sqrt{2}\left(\begin{array}{c}
-\eta^{1II}(D_0+iD_3)q 
\\
-i\eta^{1II} (D_1 + i D_2)q
\end{array}  \right)\,,
\label{transfo2} 
\end{align}
where we put $\tilde{q}$ fields to zero in the last lines.

 In the case where we have no twist, solving the BPS equation
\begin{equation}
(D_1 + i D_2)q=0
\label{bps1}
\end{equation}
would allow us to preserve $\eta^{2I}$ and $\eta^{1II}$, i.e. half of the original supersymmetry. However, when adding an angular dependency as per Eq.(\ref{twist}), the above equations show we do not have the luxury of being able to choose the relative sign, i.e. the handedness of the plane wave, the direction of its propagation. It would break all of supersymmetry if we  impose the ``wrong" choice. To preserve $\eta^{2I}$ and $\eta^{1II}$ we are forced to choose
\begin{equation}
(D_0 +  iD_3)q=0,
\end{equation}
 a mode that moves along the direction of magnetic flux (once back in Lorentzian signature). This comes at no additional cost in terms of supercharges, the object is still half-BPS. This occurs because Eq.(\ref{bps1}) is not parity-invariant,  not only does it choose a preferred axis (the unit normal axis to the $(x,u)$ plane), it also chooses a preferred direction along that axis. This parity asymmetry propagates everywhere in the BPS equations in a systematic and consistent fashion.
 
 For our approximate solution, this has the following consequence. The configuration with the ``correct" twist has energy which is close to the theoretical lower bound, given as a combination of the central charge and the (angular) momentum, which both are vectorial quantities and should point in the $\theta$ direction. Because they point in a curvilinear direction, these quantities exist only as local densities and not as total charges: there is no global $\hat{\theta}$ unit vector to express such global objects with. Nevertheless, we can express local supercharge density $\mathcal{Q}$, 4-momentum density $\mathcal{P}$  and central charge density $\mathcal{Z}$, which we do not integrate over all of space. These objects still obey the (anti-)commutation relations, locally: suppressing some space-time $\delta$-functions due to commutation,
 
 \begin{equation}
 	\lbrace \mathcal{Q}_\alpha, \bar{\mathcal{Q}}_{\dot{\alpha}}\rbrace = \sigma^\mu_{\alpha \dot{\alpha}} \left( \mathcal{P}_\mu + \mathcal{Z}_\mu\right).
 \end{equation}
 We assume invariance under rotations in the angle $\theta$. By projecting this equation on a null vector field in the $\theta$ direction, we can obtain that 
\begin{equation}
	E\geq 2\pi |\int ududx\,\, \left( \mathcal{P}_\theta + \mathcal{Z}_\theta\right) |.
\end{equation} 
 
 On the other hand, the alignment of these two vector densities has no bearing on the value of the energy:
 
 \begin{equation}
 E\geq 2\pi \int ududx\,\left(  |\mathcal{P}_\theta| + |\mathcal{Z}_\theta|\right) 
 \end{equation}
 
 The upshot is that in the case the second winding generates momentum anti-parallel to the central charge, the minimal energy configuration obtained given this requirement is very far from the theoretical minimum given by the vector sum of the two quantities, and so is far removed from being a BPS object, which we see via the SUSY algebra. This is analogous to set-up of a kink-antikink bound state, which has energy very far from the theoretical lower bound. The gap between the actual lower bound for the energy and the one dictated by the superalgebra signals gross violation of supersymmetry.
 
 Because we are using a curvilinear coordinate basis, the usual supersymmetry BPS equations should be supplemented where needed with the corresponding supergravity equations. Since we are in geometrically flat space, these simplify considerably, with one notable exception: the Killing spinor equation. It is a component of the gravitino supertransformation, thus ensuring no gravitinos are generated by curvature effects, but it is also effectively a check that parallel spinors can be found in this spacetime, in other words checking that one can define covariant spinors everywhere in space. We must solve the following equation, for $\eta$ a full Dirac spinor and $\omega_{\mu\nu\rho}$ the spin-connection of spacetime:
 
 \begin{equation}
 D_\mu \eta \, \hat{=} \, (\partial_\mu-\frac{1}{8}\omega_{\mu\nu\rho}\left[\Gamma^\nu,\Gamma^\rho\right])\eta\, = \, 0
 \end{equation}
 
 This equation obviously has a solution, as it is fully covariant and Cartesian coordinates admit constant spinors. A solution in our coordinates can be found, 
\begin{equation}
\eta=\left( \begin{array}{c}
A \epsilon^1 e^{i\frac{\theta}{2}}\\ 
B \epsilon^2 e^{-i\frac{\theta}{2}}
\end{array} \right) 
\end{equation}
where $\epsilon^{1,2}$ are Grassmann-valued Lorentz scalars and $A,B$ c-numbers. An equivalent solution is found for the lower component of the Dirac spinor. This form is entirely expected and results directly from the fact that in our coordinate system $P^3$ is an angular momentum operator in the usual Lorentz group.

As a final exercise, we can demonstrate that the mass of the settled object is a proper Hopfion, that is, one that has non-trivial Hopf index.
 
\subsection{The Hopf Invariant}\label{subs-hopf}

Such toroidal objects with two types of topological windings were observed in the form of particular field configurations of the $O(3)$ sigma model (among others), which are classified by the Hopf topological invariant
\begin{equation}
	\mathcal{H}= \frac{1}{8\pi^2}\int d^3x \epsilon^{\mu\nu\rho} A_\mu F_{\nu \rho}
\end{equation}
Such an integral also goes by the name of Chern-Simons term and has been studied extensively in the context of field theory, though usually as a term used in the construction of Lagrangians.

This topological integer can be seen to synthesise two types of winding, on very general grounds it can be expressed as the product of two other topological indices \cite{JWCW}. This is particularly clear for toroidal configurations where we can parametrise 3D space with a coordinate system that splits into one compact coordinate and an infinite plane:
 \begin{equation}
 \mathcal{H}= \frac{1}{4\pi^2} \int_{\mathbb{R}}\int_0^\infty \int_0^{2\pi}\left(  d\theta A_\theta \right) \left( ududx F_{xu}\right)  
\label{hopfinv}
 \end{equation}
The gauge field can wind around the circular direction and in the radial plane. The Hopf index is therefore an automatic indication that a given theory possesses two different types of non-trivial topological windings, and any soliton for which this quantity is non-zero can broadly be called a Hopfion.

Let us calculate (\ref{hopfinv}) for our semilocal string solution.  The component $F_{xu}$ of the field strength
is determined by the last formula in (\ref{Ansatz1}).
Moreover, the time and $\theta$ dependence of string moduli induces nonzero time and $\theta$
components of the gauge potential, see \cite{sybook}. For semilocal strings these components
were calculated in \cite{SVY} for  a non-Abelian string with $n=1$. The result obtained in \cite{SVY}
for our case of Abelian semilocal string reduces to
\begin{equation}
A_k=- i\frac{\bar{\rho}\pt_k \rho - \rho \pt_k \bar{\rho}}{u^2+ |\rho|^2}\, , \qquad k=0,3.
\label{Ak}
\end{equation}

We use Eq.(\ref{phase}) in the expression above, then, substituting $F_{xu}$ and $A_3$ into (\ref{hopfinv}) and neglecting overlap product terms, we get the following integral, whose value can be computed exactly
\begin{align}
\mathcal{H} &=  \frac{1}{\pi}k \,\int_{0}^{\infty}\int_{\mathbb{R}}ududx \left( \frac{|\rho|^4}{R((u-R)^2+x^2+|\rho|^2)^3}+\left( R\leftrightarrow -R\right) \right)  \\
&=   k\,\frac{  \left(\rho ^2+2 R^2\right)}{2 R \sqrt{\rho ^2+R^2}}\simeq k\,,
\label{hopf}
\end{align}

We can restore the dependence of $\mathcal{H}$ on the flux number $n$ considering $n$ vortex rings
located in parallel well separated planes as   in the end of Sec. 3.3. Each vortex has $\mathcal{H}\approx k$ and the Hopf
invariant, being a topological invariant, does not vary all throughout the fusion process. We conclude that
\begin{equation}
 \mathcal{H} \approx kn
\label{n-hopf}
\end{equation}
The overlap terms, terms formed by the product of two vortex profiles with different centers, can be computed also and are found to contribute terms that are  $O(\frac{\rho^2}{R^2})$, in the spirit of Section \ref{subs-error}.

Now, once the soliton has settled at its minimal length,  the form of its energy (that is, its mass) is very conspicuous: we recast Eq.(\ref{emin-n}) as
\begin{equation}
	M=\sqrt{T\left(8\pi\,n\, J_n\right) } = \sqrt{\tilde{T}}|k| |n|\sim \sqrt{\xi} \,|\mathcal{H}|
\end{equation}
where $\tilde{T}$ is the effective string tension combining the minimal string tension ($T=2\pi \xi$) times all dimensionless coefficients of the expression into a single parameter.\footnote{Recall that $J_n ={4k^2\,n\, \pi^2} \xi|\rho|^2  \log \left(\frac{1}{|\rho|\mu}\right)$.} We see that the mass of the soliton is then directly proportional to the absolute value of the Hopf invariant.


It is worth noting that, in the supersymmetric case, no absolute value is needed as both these integers have the same sign. This gives an alternate view of the case where SUSY is badly broken. Since the supercharge algebra is sensitive to the relative sign of these two windings, we can hypothesise that the theoretical lowest mass attainable, as dictated by the superalgebra, by a stable soliton is negative and therefore unphysical. It would again be the case that the actual lowest attainable mass is far removed in value from the one predicted by the superalgebra, signalling a gross violation of SUSY.

The most interesting feature of this result is that the mass is linearly dependent on the index: in the case of the $O(3)$ model, the energy functional depend non-locally on the gauge field, the fundamental degrees of freedom are scalars valued as points on a spherical target space and their energy functional satisfies a non-analytic lower bound, the Vakulenko-Kapitanskii inequality \cite{VK} 

\begin{equation}
	E\geq\left(\frac{3}{16} \right)^{\frac{3}{8}} |\mathcal{H}|^{\frac{3}{4}}
\end{equation}
This is, for the class of models the authors who proved this relation were looking at, the exact maximal lower bound for the system.

In the past, Hopfions have been constructed starting from traditional gauge theories (i.e. not $\sigma$-models), as was the case in \cite{GorskySY} and in the review \cite{Radu:2008pp}, but this was done by looking at specific configurations in the scalar sector after the gauge coupling was sent to infinity, turning the gauge field into an auxiliary field and no longer keeping it as a fundamental degree of freedom. In the process this transforms the scalar sector into a $\sigma$-model over the theory's vacuum manifold. We have been able to forgo this process here and propose a construction of a Hopfion where the topological twists are borne, either entirely or in part, directly by a fundamental gauge field in the theory.

\section*{Acknowledgments}

This work  is supported in part by DOE grant DE-SC0011842. 
The work of A.Y. was  supported by William I. Fine Theoretical Physics Institute  at the  University 
of Minnesota, by Russian Foundation for Basic Research Grant No. 18-02-00048 and by Russian State  
Grant for Scientific Schools RSGSS-657512010.2.

\end{document}